\documentclass[referee]{aa}
\input epsf.sty
\usepackage{graphicx}

\begin{document}
\bigskip
\title{Radio Emission by Particles due to Pulsar Spin}
\author{R. M. C. Thomas \and R. T. Gangadhara}
\institute{Indian Institute of Astrophysics,
        Bangalore -- 560034, India\\ mathew@iiap.res.in, ganga@iiap.res.in}

\abstract{{We present a relativistic model for the motion of charged particles
in rotating magnetic field lines projected on to a plane 
perpendicular to the rotation axis. By making an approximation that the 
projected field lines are straight, an analytical expression is obtained 
for the particle trajectory. The motive behind developing this model is 
to elucidate some of the effects of rotation in pulsar profiles.
There is a significant contribution to the curvature
of  particle trajectory due to the rotation of pulsar, which is in addition
to the inherent curvature of the field lines. The asymmetry in the observed pulse 
shapes can be explained by considering the aberration-retardation effects. The 
single sign circular polarization that has been observed in many pulsars, 
might be due to the relative orientation of sight line with respect to the 
particle trajectory plane.}
}
\authorrunning{Thomas \& Gangadhara}
\titlerunning{Radio Emission by Particles due to Pulsar Spin}
\maketitle

\section{Introduction}

 The wide diversity in the radiation characteristics of pulsars
 makes it difficult to understand fully the emission process in the
 light of models, which have been developed with some  simplifying 
 assumptions. Among
 the several emission mechanisms, curvature emission  has emerged as
 the most probable  emission mechanism (Sturrock 1971; Ruderman \&
 Sutherland 1975, hereafter RS75; Lyne \& Manchester 1988; Gil \& Snakowski 1990). 
In order to explain  the
 high brightness temperature observed in pulsars, coherent emission
 by bunched of particles has been postulated (Karpman et~al. 1975;
 RS75; Buschauer \& Benford 1977). Also, other models based on plasma effects have been
 proposed for pulsar radiation (e.g., Melrose \& Gedalin 1999; Ass\'eo
 \& Rozele 2000; Gil et~al. 2004).

            Most of these models give emphasis on explaining the high
 brightness temperature of pulsars, but the polarization is poorly
 explained.  However, the  polarization observations  such as the
 polarization angle swing favors the curvature radiation. It has been  
 considered as  a natural emission process for pulsars, though there are 
 unresolved problems
 like the  bunch formation, orthogonal polarization modes etc (e.g.,
 Stinebring et~al. 1984; Gangadhara 1997; Gil et~al. 2004).

      It is imperative to understand the influence of rotation  to
  study closely the curvature emission mechanism.  The idealized case of
  particle acceleration  has been  discussed by Machabeli \& Rogava
  (1994, hereafter MR94), who  considered particles moving
  freely  along an infinitely long, rigidly rotating straight tube
  and derived  an expression for the trajectory of a particle.
  Gangadhara \& Lesch (1997) have proposed a model for the particle
  acceleration in rotating magnetosphere in the context of active galactic
  nuclei (AGN). Reiger \& Mannhiem (2000) have also discussed the particle
  acceleration along the rotating straight magnetic field lines in
  AGN, by assuming the angular velocity of particles is same as that
  of AGN.
 
    In the case of   pulsars, Gold (1969) was the first one to
 propose a pulsar emission mechanism based on rotation.  This model
 was further  taken up by many authors and found to encounter
 difficulties  in explaining the interpulses (e.g., Sturrock 1971).
 Blaskiewicz et~al. (1991) have studied the effects of corotation velocity
 on the pulsar radio emission by  assuming  a constant emission height.
 Hibschhman \& Arons (2000)  have extended their work to include the
 first order effects to study the delays in the phase of
 polarization angle sweep due to aberration.
 Later, Peyman \& Gangadhara (2002) have improvised the model of
 Blaskiewicz et~al. (1991) by  relaxing  the assumption of constant
 emission height,   and  analyzed  the effect of rotation on the
 morphology of  pulsar profiles and polarization.

    Gangadhara (1996, hereafter G96) has derived the
 equation of motion of a charged particle in pulsar magnetosphere.  He
 has considered the straight field lines, which are projected on to a two 
 dimensional (2D) plane placed perpendicular to the rotation axis. 
 The dominant forces, which act on a particle moving along rotating the 
 field lines,  are the magnetic Lorentz force, centrifugal  force and  
 coriolis force. The rotational energy  of  the pulsar is transferred to the
 corotating plasma as it moves along the field lines.  The
 magnetic Lorentz force acts as a  constraining  force  and drags the
 plasma along the field lines.  Because of the
 inclination of magnetic axis relative to the rotation axis,
 corotating plasma tends to rotate  with an angular velocity
 less than that of pulsar  on some  field lines.  The difference in
 the angular velocities of particle  and pulsar had been already
 pointed out in RS75.

    In   the present  work,   as  a  follow up of G96,
 we consider  the same 2D geometry and analyze the dynamics of a
 charged particle.  Since the field line curvature radii of open field
 lines  are comparable   to the light cylinder radius,  over  a
 significant radial distance, we can approximate them to be straight lines.
  
    In super strong magnetic fields, the drift velocity becomes negligible  compared
to the velocity parallel to the field lines. The Larmour radius of gyration 
becomes quite small, and hence particles almost
stay on the same field lines all along their trajectories. This
motion is considered as the bead-on-wire approximation. The particles
are accelerated because of the unbalanced centrifugal force, and thus
extract the rotational energy of the pulsar. The single particle emission is 
considered in this model and plan to consider the collective effects in our
follow up works. We take into
consideration of the non-uniform angular velocity of  particles, which
can be less than the pulsar angular velocity on some field lines which
are inclined with respect to the meridional plane.
  Since the particle trajectories
are found to be curved, we estimate the curvature emission and analyze the
effects of rotation  on the  radiation characteristics.  In
\S~\ref{sec_dynamics} and \S~\ref{sec_trajectory}, we   solve the
equation of motion of a relativistic charged particle  and find its
trajectory. We compute the characteristic frequency of  curvature
radiation in \S~\ref{sec_radiation}.  In \S~\ref{sec_stokes}  we
estimate  the polarization parameters and plot them with respect to different
parameters.
 
\section{Charged particle dynamics }\label{sec_dynamics}

\begin{figure}
\begin{center}
\epsfxsize=12cm \epsfysize=0cm \rotatebox{0}{\epsfbox{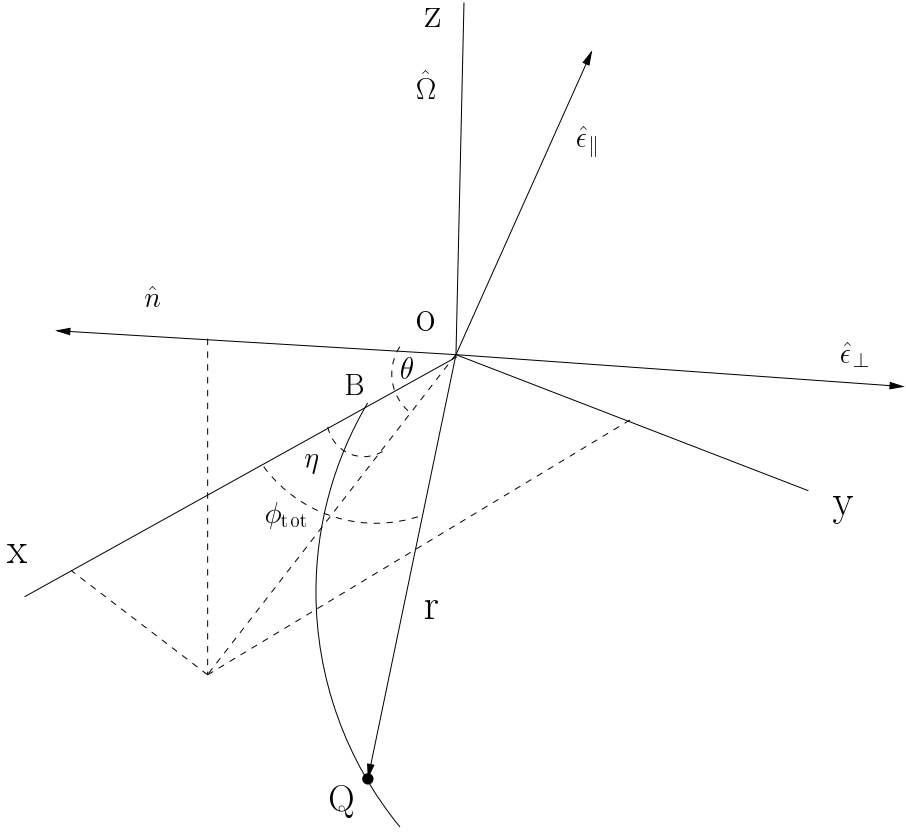}}
\caption{The coordinate system in which the
    particle motion is considered. The curve BQ represents the
    particle trajectory in the $x$-$y$ plane.
\label{fig(caxis)}}
\end{center}
\end{figure}

 We assume that the dipolar magnetic  field lines are projected onto a
plane perpendicular to the rotation axis. Consider   an inertial
Cartesian coordinate  system as shown in Fig.~\ref{fig(caxis)}, where
the 'z' axis is parallel to the rotation axis ($\hat{\Omega}$) of
pulsar.  The projected  magnetic axis on the x-y plane coincides with
the $x$-axis at time $t =0$.  The equation of motion for a charged
particle moving along a rotating magnetic field line is given by
(G96),
\begin{equation}\label{eqmotion}
\frac{d}{dt}\left (m\frac{d r}{dt}\right) = m {\Omega}^{*2}r,
\end{equation}
where $\rm{ m=m_{0}{\gamma}} $ is the relativistic mass, $\gamma$  the
Lorentz factor, $m_0$ the rest mass, $\Omega^{*}$  the  angular velocity
and $r$  the radial position of a  particle.

   Let $ V_r=dr/dt\,$ and $V_\phi=r\Omega^{*}$ be the components of
 particle velocity, then
\begin{equation}\label{eqbeta}
  \vec{\beta}=\frac{1}{c}(V_r\hat{e}_r + {V}_\phi\hat{e}_\phi)  ,
 \end{equation}
   where $c$ is the speed of light. We define the unit vectors in the
radial and azimuthal directions  as
\begin{eqnarray}
\hat{e}_r &=& (\cos\phi, \, \sin\phi,\,  0)\, ,\\
\hat{e}_\phi &=& ( -\sin \phi,\, \cos\phi,\, 0)\, ,
\end{eqnarray}
where $\phi$ is the angle between the radial vector to particle and  
the $x$-axis.
 Then the Lorentz factor of  particle is given by
 \begin{equation}\label{gamma}
 \gamma  = \frac{1}{\sqrt{1-\beta^2}} = \left[1-
 \left(\frac{1}{c}\frac{dr}{dt}\right)^2 -
 \left(\frac{r\Omega^{*}}{c}\right)^2\right]^{-1/2}.
  \end{equation}

\begin{figure}
\begin{center}
\epsfxsize=9cm \epsfysize=0cm \rotatebox{0}{\epsfbox{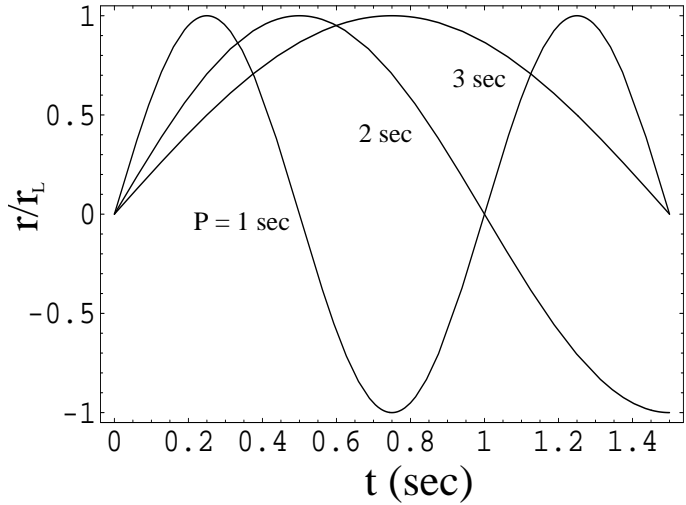}}
\caption{Radial position of the  particle  as a
function of time. Used $\gamma_0=100,$ $d_0=10^6$ cm and
$\theta_0 = 90^{\circ}$.
\label{radial_distance}}
\end{center}
\end{figure}

    Consider a particle injected at the point B onto a magnetic field
line  which is inclined by an  angle $\rm{\phi_p}$ with respect to the 
$x$-axis at time $t=0.$ Let  $d_0 = \rm{OB}$ be the distance
between B and the rotation axis.
The effective angular velocity (G96) of a particle is given by
\begin{eqnarray}\label{Omega*}
   \Omega^{*}&=& \Omega\left[\frac{r^2-d_0^2\cos^2\theta_0 -d_0
\sin\theta_0\sqrt{r^2-{d_0}^2\cos^2\theta_0}}
{r\left(\sqrt{r^2-{d_0}^2\cos^2\theta_0}
- d_0\sin\theta_0\right)}\right]\nonumber\\
&=& \Omega\sqrt{1-\frac{b^2}{r^2}},
\end{eqnarray}
where $\Omega$ is the angular velocity of  pulsar, $b=d_0
\rm{cos\theta_0}  $ and $\theta_0=(\pi /2)-\phi_p$ is the angle
between the field line tangent and $\hat{e}_{\phi}$ at B.  Using the 
relation for  $\rm{\Omega^*},$  we can write $\gamma $ as
\begin{equation}\label{eqgamma}
\rm{\gamma=\left [1+D^2 - \left(\frac{1}{c}\frac{dr}{dt}\right)^2-
\left(\frac{r\Omega}{c}\right)^2\right]^{-1/2}},
\end{equation} 
where $D= \Omega d_{0}\cos\theta_{0}/c.$ Thus, using the expression for  
$\gamma,$ we rewrite Eq.~({\ref{eqmotion}}):
\begin{eqnarray}\label{eqmotion_1}
  \gamma \frac{d^2r}{dt^2} +\frac{d\gamma}{dt}\frac{dr}{dt}=
\Omega^{2}\left(1-\frac{b^2}{r^2}\right)\gamma\, r~.
\end{eqnarray} 
By multiplying Eq.~(\ref{eqmotion_1}) by  $r/(\gamma c^2),$ and
defining   a dimensionless  variable
\begin{equation}\label{eqtrans}
    s=\frac{\Omega}{c} \frac{r}{\sqrt{1+D^2}} \ ,
\end{equation}
   we obtain 
\begin{equation}\label{eqcomp}
    s\frac{d^2s}{dt^2}+   \frac{\left[2s^2 -D^2/(1+D^2)\right]}{1-s^2}
    {\left(\frac{ds}{dt}\right)}^2 - s^2\Omega^{2}
    +\Omega^{2}\frac{D^2}{1+D^2} =0 .
\end{equation}
 Since $\theta_0$ is close to $\pi/2$ for the field lines, which are
 close to the $x$-axis,  we find  $D^2 \ll s^2$ for $  d_0 < r$.
 Therefore, we reduce Eq.~({\ref{eqcomp}}) by dropping the  terms
 containing $ D^2/1+D^2,$ and obtain
 \begin{equation}\label{eqnew}
  \frac{d^2s}{dt^2} + \frac{2s}{1-s^2}\left(\frac{ds}{dt}\right)^{2}
 -s\,\Omega^{2}=0 \,.
 \end{equation}
To find the solution of Eq.~(\ref{eqnew}), we follow the method proposed
by Zwillinger (1989). By choosing $f=(ds/dt)^2$, we
can be reduce it to the following form:
\begin{equation}\label{eqnew1}
  \frac{df}{ds}+f\frac{4s}{1-s^2}=2s\, \Omega^2\, .
 \end{equation}
Its solution is given by
  \begin{equation}\label{eqmachabeli}
  f=\Omega^{2}(1-s^2)+ C(1-s^2)^2,
  \end{equation}
where $C$ is the  integration  constant. To find $C$ we use the 
initial condition that at t=0, it follows from Eq.~({\ref{eqtrans}}) that
   $s_{0}=s(0)=r_0\Omega /(c \sqrt{1+D^2})$
 \quad and
  $\dot{s}_0=ds/dt|_{t=0}=v_0\Omega/(c \sqrt{1+D^2}),$
 where $r_0$ and $v_0$ are the initial position and velocity of
 particle.  Therefore, we obtain $ C=-\Omega^2 k^2 $ and
   \begin{equation}\label{eqk}
  k^2=\frac{1}{1-s^2_0}\left[1-
       \frac{{\dot{s_0}}^2}{(1-s^2_0)\Omega^2}\right].
   \end{equation}
Hence  Eq.~({\ref{eqmachabeli}}) reduces  to the following form
\begin{equation}\label{eqvarf}
      \frac{ds}{dt}=\Omega\sqrt{(1-s^2)-k^2(1-s^2)^2}
\end{equation}
whose solution is given by
\begin{equation}\label{eqjacob}
  s = \rm{cn}({\lambda}-{\Omega} t),
\end{equation}
where $\rm{cn}(z) $ is the Jacobian Elliptical cosine function
      (Abramowicz \& Stegun 1972), and
\begin{eqnarray}\label{eqz}
       z = \int\limits_{0}^{\rm{sn(z)}}\frac{dw}{\sqrt{(1-w^2)(1-k^2w^2)}}\, ,
\end{eqnarray}
\begin{eqnarray}\label{eqlambda}
  \lambda=\int\limits_{0}^{\phi_0}\frac{d\zeta}{\sqrt{1-k^2{\sin^2{\zeta}} }}\, ,
\end{eqnarray}
 \begin{eqnarray}\label{eqphi}
      \phi_0=\arccos\left({\frac{r_0\Omega}{c}}\right)\ .
\end{eqnarray}
Using the expression for $s$ given by Eq.~({\ref{eqtrans}}), we  find the
radial position of the  particle:
\begin{eqnarray}\label{eqrad}
   r = \frac{c\sqrt{1+D^2}}{\Omega} \rm{cn}(\lambda-\Omega t)\ .
\end{eqnarray}
  The radial position of the particle according to
 Eq.~({\ref{eqrad}}) as a function of time  is
plotted in  Fig.~\ref{radial_distance}. It shows that the particle
position increases with time and reaches a maximum at the distance of 
light cylinder
radius $r_L=Pc/2\pi,$ where P is the pulsar period.  Next,  the
particle returns back to origin, due to the reversal of the
centrifugal force. This  type of oscillatory motion  of a  particle
in  an infinitely long, straight and rigidly rotating tube  has been
discussed by MR94.

Though we have extended the calculation of $r$ of a single particle
all the way up to light cylinder, it may not be realistic in the case
of  plasma motion. Near the light cylinder, plasma inertia causes the
field  lines to sweep back and break down of the rigid body motion.
The  Lorentz factor of a particle, which follows from Eqs.~({\ref{eqgamma}}),
({\ref{eqvarf}}) and ({\ref{eqjacob}}), is  given by   
\begin{equation}
          \gamma=\frac{1} {k \sqrt{\left(1+D^2\right)}
    {\rm sn}^{2}({\lambda}-{\Omega t}) }\ .
\end{equation}
    
\section{Particle trajectory and its radius of curvature}\label{sec_trajectory}

\begin{figure}
\begin{center}
\epsfxsize=6 cm \epsfysize=0cm \rotatebox{0}{\epsfbox{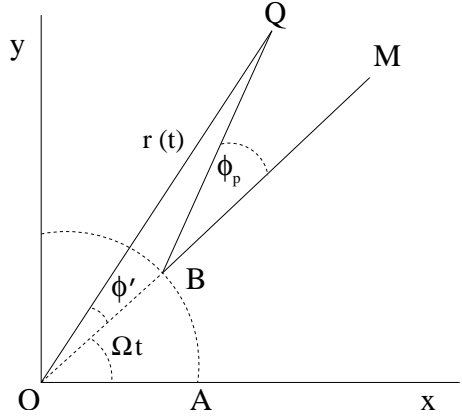}}
\caption[]{The geometry of motion of a particle along a rotating field line BQ.  
The angles are $\angle {\rm XOM} = \Omega t,$  $\angle {\rm MBQ} = \phi_{\rm p}$ and 
$\angle {\rm MOQ}=\phi ',$ and the radius OA=OB=$d_0.$
\label{phitot}}
\end{center}
\end{figure}

 In Fig.~{\ref{phitot}} we consider a particle moving along the
field line BQ. The point A represents the particle injection point at time $t=0$ that
is at a distance $d_0$ from the rotation axis. The particle  
co-ordinates can be defined as
\begin{equation}\label{eq_xy}
 (x,\, y)=r(t)\left(\cos{\phi_{\rm{tot}}},\, \sin{\phi_{\rm{tot}}}\right) ,
\end{equation}
where $\phi_{\rm{tot}}$ is the angle between the  radial vector to the
particle and the  $x$-axis.  Based on Fig.~{\ref{phitot}},  we
define
 \begin{equation}\label{eq_phi_tot}
   \phi_{\rm{tot}} (t)= \Omega t \pm \phi'(t) \, ,
\end{equation}
    where
\begin{equation}\label{eq_phi_dash}
     \phi ' (t)=  \cos^{-1}\left(\cos\phi_{\rm{p}} \sqrt{1-\
       \frac{{d_0}^2}{r^2}\sin^{2} \phi_{\rm{p}}} \quad+\quad
       \frac{d_0}{r}\sin^{2}\phi_{\rm{p}}\right) .
\end{equation}
 For $d_0\ll r,$ we find
\begin{equation}\label{eq_phi_approx}
     | \phi' (t)| \simeq | \phi_{\rm{p}} |
 \end{equation}

\begin{figure}
\begin{center}
\epsfxsize=10cm \epsfysize=0cm \rotatebox{0}{\epsfbox{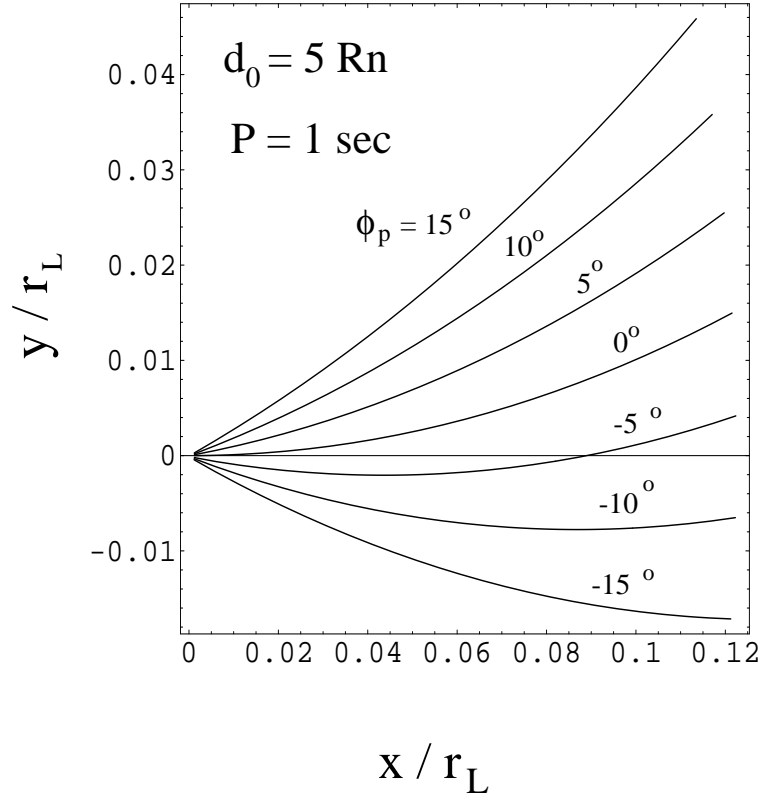}}
\caption{Particle  trajectories during the time
 interval $0 \leq t \leq 0.02$~sec  in laboratory frame. We considered the 
 field lines which lie  in the range of $-15^o \leq \phi_p \leq 15^o$ with
 an interval of  $5^o.$ Assumed neutron star radius $R_{\rm n} \approx 10$~Km.
\label{trajectory}}
\end{center}
\end{figure}

The $\pm $ signs  in Eq.~(\ref{eq_phi_tot})  corresponds to the sign
of the angle $\phi_{\rm{p}}.$ In Fig.~{\ref{trajectory}} we have
plotted the trajectories   of the particles moving along different
magnetic field lines, which are  marked with  $\phi_p.$ It shows that the
trajectories are curved towards the direction of rotation of pulsar.
The  particles moving in those trajectories are accelerated, and hence  
they emit curvature radiation. The curvature radii of those
trajectories  slightly  differ from one another, as the particle
angular  $\Omega^*$ is different for each  field line.

To derive the curvature radii of the particle trajectory, we
approximate $\rm cn(\lambda-\Omega t)$  and $r(t)$ using the
formalism given by Pearson (1974):
\begin{equation}\label{eqra}
cn(z,k)=\cos{z}+\frac{k^2}{4}(z-\sin{z}\cos{z})\sin{z}+O(k^4).
\end{equation}
In the limit of $t\ll 1$ and $k\ll 1$,  the series expansion 
of $r(t)$ is given by
\begin{eqnarray}\label{eqserex}
       r(t)=a_0+a_1t+a_2t^2+a_3t^3+a_4t^4.....
\end{eqnarray} 
where $a_0,\, a_1,\, a_2,\, a_3,\, a_4...$ are the expansion
coefficients:
\begin{eqnarray}\label{eq_coefficients}
a_0 & = & \frac{c\,{\sqrt{1 + D^2}}}{\Omega}\,\Big[ \cos \lambda  +
      \frac{{{k}}^2\,\sin \lambda \,
         ( {\lambda } -
           \cos \lambda \,\sin \lambda  ) }{4}\Big] ~,\\
 a_1 & = & \frac{c\,{\sqrt{1 + D^2}}}{16}\,
\Big[  2\, \sin \lambda \{ 8 - k^2 + 3\,k^2\,\cos (2\,\lambda ) \} \,
         -4\,k^2\,\lambda \,\cos \lambda   \Big] ~,\\
 a_2 &=& \frac{- c\,{\sqrt{1 + D^2}}\,\Omega \,}{32}
      \Big[ \left( 16 - 9\,k^2 \right) \,\cos \lambda  +
        k^2\,\left( 9\,\cos (3\,\lambda ) + 4\,\lambda \,\sin \lambda  \right)
\Big] ~,\\
  a_3 &=& \frac{c\,{\sqrt{1 + D^2}}\,{\Omega }^2\,}{96}
    \left[   k^2\,
       \left\{ 4\,\lambda \,\cos \lambda  + 13\,\sin \lambda
- 27\,\sin (3\,\lambda ) \right\}  -16\,\sin \lambda
          \right]  ~,\\
 a_4 & =& \frac{c\,{\sqrt{1 + D^2}}\,{\Omega }^3\,}{384}
    \left[ ( 16 - 17\,k^2 ) \,\cos \lambda  +
      k^2\,\{ 81\,\cos (3\,\lambda ) +
         4\,\lambda \,\sin \lambda  \}  \right]~.
\end{eqnarray}

For $\rm{v_0\approx c},$ Eq.~(\ref{eqk}) implies $k\approx 0.$ Therefore, using 
Eqs.~({\ref{eqjacob}}), ({\ref{eqz}}), ({\ref{eqlambda}})
and ({\ref{eqphi}}), we find $\lambda=\pi/2$ and $\rm{sn}~z=\rm{sin}~z.$ Thus, 
we have
\begin{eqnarray}
       r(t)\approx \frac{c\,{\sqrt{1 + D^2}}}{\Omega}\,\sin{(\Omega t)} \, .
\end{eqnarray} 
      Using the expression for $r(t)$ and Eqs.~(\ref{eq_xy}) and
 (\ref{eq_phi_approx}), we find  the curvature radius of particle trajectory:
\begin{eqnarray}\label{eq_rho}
\rho & = & \frac{[(dx/dt)^2+(dy/dt)^2]^{3/2}} {(dx/dt)
              (d^2y/dt^2) -(dy/dt)(d^2x/ dt^2)}\nonumber \\  
     &\approx & \frac{1}{2} r_L \sqrt{1+D^2}.
\end{eqnarray}
        It shows that the curvature radius of a particle trajectory is
approximately $r_L/2$ for $\theta_0=\pi/2.$ However, for other
values of $\theta_0,$ we find $\rho$ becomes slightly larger than $r_L/2.$
Note that these values of $\rho$ are comparable to the intrinsic curvature 
radii of dipolar field lines in the emission region given by Gangadhara
(2004). We find, in the conal emission regions, the intrinsic 
curvature radii of field lines
are comparable to the curvature radius induced by rotation into the particle
trajectory. If the core emission is believed to come from the field lines 
which are close to the magnetic axis then it becomes difficult explain the
core emission due to the intrinsic curvature of field lines. This is because 
the field lines which are very close to the magnetic axis have very large
curvature radii, and for the magnetic axis it is infinity. Therefore, in
the absence of rotation, we can not expect any significant curvature emission
from the particles or plasma moving along the field lines which are close to 
the magnetic axis.  However, from the observations we do see many pulsars 
emitting strong cores.  In our model we show that if we consider rotation
of particles, the core emission can be explained. Because the
rotation induces significant curvature into the trajectory
of particles which move along the field lines
that are  close to the magnetic axis.

\section{Radiation emitted by a particle}\label{sec_radiation} 
   
\begin{figure}
\begin{center}
\epsfxsize=10cm \epsfysize=0cm \rotatebox{0}{\epsfbox{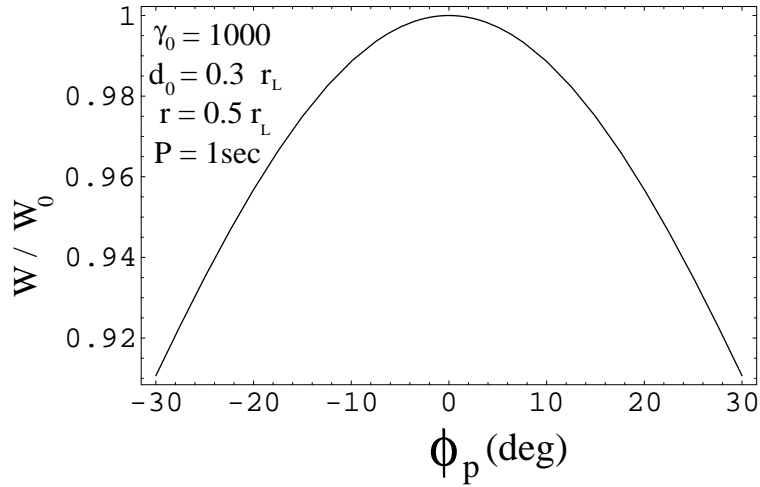}}
\caption[]{The power emitted by a particle as a function of $\phi_p.$  
The parameter $W_0 = 25.6  q^2 \Omega^2/(3c {k_0}^4),$ and $k_0=k$ at $D=0.$
\label{power_phi}}
\end{center}
\end{figure}

       When the particles move along  curved trajectories, they emit
curvature radiation. The characteristic frequency of the curvature  
radiation is given by (RS75)
\begin{eqnarray}\label{eq_ch_frequency}
 \omega_c & = &\frac{3}{2}\gamma^3\left(\frac{c}{\rho}\right)\nonumber \\  
          &\approx & 3\, \Omega \frac{1+D^2} {k^3\left[ 1+D^2-(r/r_L)^2\right]^3}\, .
\end{eqnarray}
 By having known the $\gamma$ and $\rho$ from  Eqs.~(\ref{eqgamma})
and  (\ref{eq_rho}), we can estimate $\omega_c.$ 
The total power emitted by a particle is given by
\begin{eqnarray}\label{eq_power}
 W & = & \quad  \frac{2}{3} \frac{q^2}{c} \, \,\gamma^4
         \left(\frac{c}{\rho}\right)^2\nonumber \\ \nonumber \\ 
   & \approx & \frac{8\, q^2}{3\, c}\,    \frac{ \Omega^2 (1+D^2)}{k^4
               [1+D^2-(r/r_L)^2 ]^4} \, ,
\end{eqnarray}
where $q$ is the particle charge. 
In Fig.~\ref{power_phi}, we have plotted $W$ as a function of $\phi_p$ 
at $r =0.5\, r_L.$ We find maximum power
is emitted by the particles which move along the field line  with
$\phi_p=0$ as $D=0,$ compared to those on the other field lines $(\phi_p\neq 0).$

\section{Polarization of Radiation}\label{sec_stokes}
           
The radiation electric field is given by (e.g., Jackson 1972; Gangadhara 1997)
\begin{equation}\label{eqefield}
{\bf E}(\omega)=C_f\int\limits_{-\infty}^{+\infty} \hat{n}\times(\hat{n} 
      \times\ {\vec{\bf\beta}})\,\rm{exp}\{\,i\, \omega(t-\hat{n}.{\bf{\vec r}}/c )\}\, dt\, ,
\end{equation}
where $ C_f=-i\omega\ q\, e^{i\omega S_0/c}/\sqrt{2\pi}S_0 c, $\, 
$S_0$ is the distance from  the origin to observer, $\omega$  the radiation 
frequency and $\hat n$ the sight line.  To solve the integral, we shall express 
$\hat{n}\times(\hat{n} \times \bf{\vec{\beta}})$ (see, Appendix-A) and the   
argument of exponential as series expansions in time $t.$ 
Consider  the sight line  which  makes an angle $\theta$ with 
the 2D plane, and  $\eta$ with the $x-$axis:
\begin{equation}\label{eq_l_sight}
     \hat{n}=(\cos\theta \cos\eta,\, \cos{\theta}\sin{\eta},\, \sin{\theta}) .
\end{equation}
To describe the polarization state of the emitted radiation, we define  orthogonal 
unit vectors (see, Fig.~\ref{fig(caxis)}):
\begin{eqnarray}\label{eq_set_vec}
\hat{\epsilon}_\parallel & = & ( -\sin\theta \cos\eta,\, -\sin\theta\sin\eta ,\, \cos\theta)\, , \\
\hat{\epsilon}_\perp & = & (-\sin\eta,\, \cos\eta,\, 0)\,. 
\end {eqnarray}
 The unit vectors ($\hat{n},\,
  \hat{\epsilon}_{\parallel},\,\hat{\epsilon}_{\perp}$) form an
  orthogonal triad:
\begin{equation}\label{eq_identity}
 \hat{n}\times\hat{\epsilon}_{\perp}=\hat{\epsilon}_{\parallel}\ .
 \end{equation}

Let $t_0 $ be  the time at which $\vec{\beta}$ aligns with $\hat{n},$
and the observer receives radiation. We transform the time
variable t to $t+t_{0}$ such that $\rm{\Omega t_{0}}$ stands for an
initial phase.  Thus, we find
\begin{eqnarray}\label{eq_rt0}
      r(t+t_0) = a'_0 + a'_1 t + a'_2 t^2 + a'_3 t^3 + a'_4 t^4\, ...
\end{eqnarray}
based on  Eqs.~(\ref{eqra}) and (\ref{eqserex}).  The expansion coefficients 
$a'_0,\, a'_1,\, a'_2,\, a'_3,\, a'_4\, ...$ are same as  $a_0,\, a_1,\, 
a_2,\, ... $  except $\lambda$ replaced by $\lambda-\Omega t_0.$ 

Using Eqs.~(\ref{eqrad}) and~(\ref{eq_xy}), we find the series expansion of 
the exponential argument in Eq.~(\ref{eqefield}) and keep the terms up to the 
order of $t^3$:
\begin{eqnarray}\label{eq_N_0123}
\omega\left[(t+t_0)-\frac{\hat{n}.\vec{r}}{c}\right] 
& = & \omega\left[t+t_0 - \frac{ r }{c}\cos\theta \left(\cos\eta\,
      \cos\phi_{\rm{tot}} + \sin\eta\,  \sin\phi_{\rm{tot}}\right)\right]\, ,  \nonumber\\ 
& = & N_0 + N_1 t + N_2 t^2 + N_3 t^3\, ....
\end{eqnarray}
 
The coefficients $N_0,\, N_1,\, N_2$ and $N_3$ are given in the Appendix-C.
The series expansion
of exponential argument is converging, and it is quite obvious
from Eqs.~(\ref{eqrad})\, and\,  (\ref{eqra}). In the limit  of
$k\approx 0,$ the series expansion of $r$ behaves like  the
trigonometric sine function. Since the angular width of emission beam is 
$\approx 2/\gamma,$ the time taken by the particle to cross the angular 
width of the order of emission  beam is  $\approx 2\rho/c\gamma.$ Thus the  
truncation of higher order terms  introduce a  negligible error in our
calculations. Since we intend to  reduce the integral in Eq.~(\ref{eqefield}) 
to a known form, we limit the series expansion terms up to the order of $t^3$.
  
Using the transformation given by Buschauer \& Benford (1976), we find the 
electric field components  (see, Appendix-B):
\begin{eqnarray}
E_{\parallel} & = & \frac{1}{c}\left(V_{\parallel 0}B_{0}+V_{\parallel 1} B_1
                     + V_{\parallel 2}B_2\right) \, C_f\, e^{iN_0}\,
                      \sin\theta \, , \\ 
    E_{\perp} & = & \frac{1}{c}\left(V_{\perp 0}B_0 +V_{\perp 1}B_1+V_{\perp 2}B_2
                    \right)\ C_f\,e^{iN_0}\, .
\end {eqnarray}            
Next, we define the Stokes parameters as
\begin{eqnarray}
I & = & E_{\parallel} E_{\parallel}^{*} + E_{\perp} E_{\perp}^{*} \, ,\nonumber \\
Q & = & E_{\parallel} E_{\parallel}^{*} - E_{\perp} E_{\perp}^{*} \, ,\nonumber \\
U & = & 2\Re\left(E_{\parallel}^{*} E_{\perp}\right) \, ,\nonumber \\
V & = & 2\Im\left(E_{\parallel}^{*} E_{\perp}\right)\, .
\end{eqnarray}
 The linear polarization is given by
\begin{equation} 
L = \sqrt{Q^2 +U^2} \, .
\end{equation}
 In observations, pulsar polarization is normally expressed in terms of
$L$ and $V$. So, in our model, we call $L$ and $V$ as polarization parameters.

\subsection{Polarization parameters of radiation emitted by many  particles}\label{sec_stokes_group}

 We consider a set of field lines on the 2D plane, and estimate the
 total emission  by particles accelerated  along them.  During pulsar
 rotation, the sight line stays at a particular $\theta$ with respect
 to the 2D plane. Since the emission from each particle is
 relativistically beamed in the direction of velocity $\vec{\beta},$
 the observer tends to receive the  radiation  from all those particles,
 for which $\vec \beta$ falls with in the angular width  of  $\pm
 1/\gamma$ with respect to   $\hat n.$
         
     First we estimate the polarization parameters of the radiation
 emitted by a single particle  at the instant $t_0\leq t_{\rm max}.$ The 
 instant $t_0$ is the time at which $\hat{n}. \vec{\beta} =1 $ for a 
 given initial $\phi_p.$ As the rotation progresses, new $t_0$ is 
 computed for the advanced rotation phase by again solving 
 $\hat{n}. \vec{\beta} =1,$ and computed the polarization parameters. This 
 procedure is continued till $t_0 \approx t_{\rm max},$ where $t_{\rm max}$ is 
 the time at which the particle goes out of radio emission zone 
 ($d_0\leq r \leq 3\times 10^{3}\,$ Km). Since the radiation is emitted 
 over a range of $r,$ due to the aberration and retardation the radiation
 beam gets shifted to the leading side of the pulse. The role of 
 retardation and  aberration phase shifts has been discussed by e.g., 
 Phillips (1992) and Gangadhara \& Gupta (2001).
     
   In order to compute the  total   polarization parameters with respect to
the  rotation phase, first we sort the  polarization parameters due to  single
particles into groups of phase bins  and add them. In the following steps, 
we give the details of the procedure:
\begin{enumerate}
\item  Fixed the observer's sight line at a  specific $\theta$
       with respect  to the 2D plane.
\item  Selected a set of field lines in the range of   $ -5^\circ \leq
       \phi_p \leq 5^\circ $ with  a successive line  spacing of
       $0.1^\circ.$
\item  Solved $\hat n.\hat \beta= 1$ to find $t_0$  at the point of
       emission  on the trajectory corresponding to each field line, and
       estimated the polarization parameters at  those points.
\item  The retardation phase shift $\Omega(t_{\rm{\rm max}}-t_0)$ is subtracted
       from  $\eta$ assigned for each of the emission beam, and estimated the
       effective rotation phase.
\item  Next, the sight line is rotated by $0.1^\circ$  to a new phase,
       and repeated the procedure (1-4) over the range of $-12^\circ \leq
       \eta\leq 12^\circ.$
\item Finally,  the array of polarization parameters  are sorted into
      groups of phase bins, and  added to  get the total pulse profile.
\end{enumerate} 
  
\begin{figure}
\begin{center}
\epsfxsize=15cm \epsfysize=0cm
\rotatebox{0}{\epsfbox{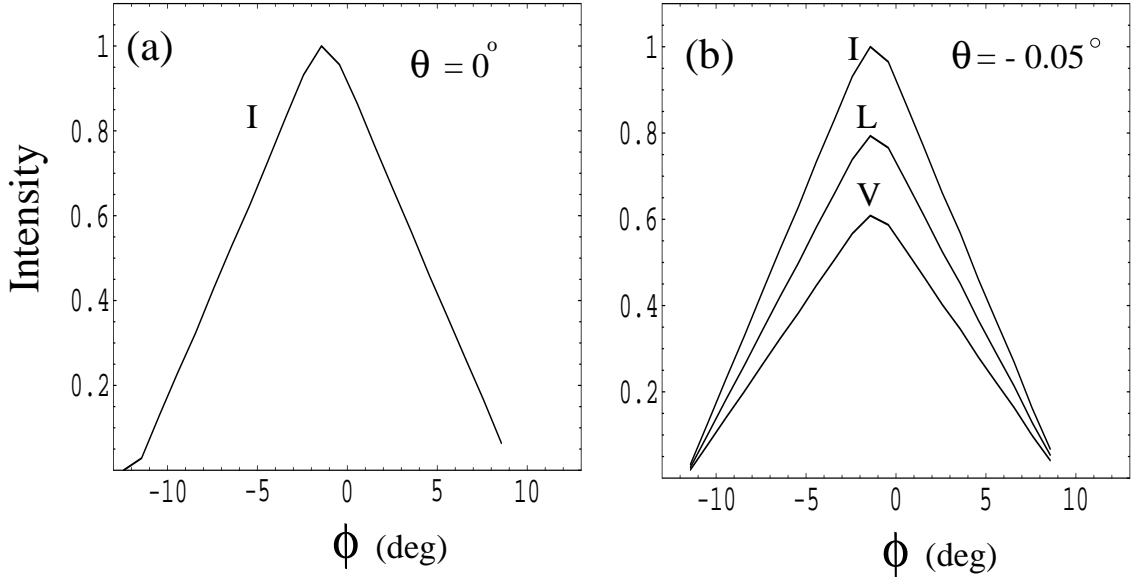}}
\caption[]{The simulated profiles: panel (a) for $\theta = \rm{0^{\circ}}$ and 
 panel (b) for $\theta=-0.05^{\circ}.$ The parameter $\phi$ is the rotation phase. 
 Used $\gamma_0 = 100$ and $d_0 = 10$~Km. \label{fig(intensity)}}
\end{center}
\end{figure}

In  Fig.~\ref{fig(intensity)}, we have given the total polarization parameters 
computed from the emissions by many particles as functions of rotation phase. 
In panel (a) we have plotted the profile that is obtained when the sight line 
lies in the 2D plane,  and panel (b) for the case when  the sight line is 
inclined by  $-0.05^{\circ}.$ The profiles indicate that the peak emissions are 
shifted to the earlier phase as a consequence of the aberration-retardation effect. 

 \section{Discussion}

 For simplicity we considered the  dipole field lines projected on
 to a 2D plane. The relativistic particles are assumed to be streaming outward
 along such a field configuration with an initial Lorentz factor in the range of
 $10^2\leq\gamma_0\leq 10^3.$ Since our aim is to understand the rotation effects on
 particle dynamics and pulse profile, we consider  the single
 particle emission, and leave  the collective plasma emissions to
 latter works. We are interested in the region which extends from a few stellar
 radii to a radial distance well within the light cylinder, where the
 radio emission is expected to occur and the  bead-on-wire approximation holds.
 Though we have approximated the particle motion to 2D, we are able to 
 investigate the influence of rotation on pulsar profiles. Our model is 
 more relevant for the
 cases  where the inclination angle $\alpha$ of the magnetic axis
 relative to rotation axis is large enough. In such  cases, the
 projected field lines may be approximated to be straight lines over a
 significant radial distance.   We have derived an expression   for
 the radial position of a particle  (Eq.~\ref{eqrad}), which   shows
 an oscillatory behavior, as shown in Fig.~\ref{radial_distance}.  A
 similar case of  particle motion    in an infinitely long, straight
 and  rigidly   rotating tube, has been  discussed by considering  a
 gedanken experiment by MR94.  They have shown
 that due to the  centrifugal force reversal,  particle  returns back
 to the rotation axis after reaching a   maximum  distance at which  the
 rotation velocity reaches the speed of light. It turns out to be an
 oscillatory motion in the radial direction.  
 Gangadhara (1996) has shown  that the particle
 angular velocity cannot be  same as the  field line angular velocity
 if the magnetic axis is inclined with respect to the rotation axis.  We
 considered this effect in our treatment of particle motion, and
 found the  particle trajectories  and their   curvature radii vary
 with field line orientation.

 Since the magnetic field is very strong, Larmour radius of gyration and drift velocity 
becomes very small. So, the particles are assumed 
 to follow the same set of field lines all along their trajectories. 
 In the case of  single particle dynamics, the magnetic field dominates
 and hence the rigid body motion may be extended all the way up to
 the light cylinder. But in reality plasma corotates with the  neutron
 star, and we must to take into account of the plasma inertia  in the
 region close to the light cylinder.  Therefore, it is possible that
 the magnetic field lines will sweep  back, and can lead to the generation of
 toroidal magnetic field. Hence the oscillatory motion that our
 solution predicts, can not be achieved in a real physical
 situation like pulsars. So, the particle which reaches the vicinity
 of the light cylinder can not come back, but escape from the
 magnetosphere as a pulsar wind.  
 
 We find the energy of particle increases due to the centrifugal
 force, as indicated by the Eq.~(\ref{eqgamma}) for Lorentz
 factor $\gamma .$ By this way the rotational energy of neutron star gets
 transferred to the particles via the magnetic field lines.

 We find  the radius of curvature of particle trajectory is
 approximately $r_L/2,$ which is comparable to the inherent radius of
 curvature of dipolar field lines (Gangadhara 2004).   So, we believe the
 curvature emission due to the rotational motion of particles should
 be comparable to the actual curvature emission in corotating frame.
 Both the Lorentz factor and the characteristic  frequency reach the maxima 
 at the region close to light cylinder. Hence our model indicates that the 
 high frequency radiation (e.g., X-ray, $\gamma$-ray) may be emitted in the 
 regions close to light cylinder.

In a later work,  Rogava et~al. (2003) have shown that if a particle freely 
moves along a tube with an
arbitrary curvature, the centrifugal force does  not reverse
always. They have showed that the particles move in the tube with 
a variable angular velocity. This supports our result that the particles 
angular velocity on some field lines differs from that of pulsar.
That is the particles moving along the field lines with $\phi_p=0$ rotate with
the angular velocity which is same as the pulsar angular velocity.
But those moving along other field lines, for which $\phi_p\neq 0,$
rotate with the angular velocity which is smaller than the pulsar
angular velocity. The particles moving along the field lines with
$\phi_p\sim 0,$ tend to emit more power than those moving along other
field lines, and hence the profile in Fig.~\ref{power_phi} shows a peak at 
$\phi_p\sim 0.$ Also, it is evident from observations that the peak of pulsar 
profiles (core) is, probably, emitted from the  field lines with $\phi_p\sim 0.$

       By taking into account of aberration-retardation,  we
have reproduced a simulated a pulse profile (Fig.~\ref{fig(intensity)}) 
by adding the  radiation emitted by particles accelerated on a set of 
field lines. The sign of $\phi$ has been flipped to match with the  
phase sign convention followed in pulsar profiles. The roughness
in the curves of Fig.~\ref{fig(intensity)} are due to the increments
of $0.1^\circ$ in $\phi_p$ and $\eta.$ This choice was made based on the
limitation in computing time. However, the smoother profiles can always 
be generated by choosing smaller increments and opting for longer computing
time. Since we consider an uniform plasma flow along the field lines, our 
profiles do not have subpulse components.
 
     Our model shows the effects such as the aberration and retardation
makes the pulse profiles to become asymmetric about the pulse center.
This phenomenon has been observed in most of the pulsar profiles
(e.g., Gangadhara \& Gupta 2001; Gupta \& Gangadhara 2003).

In our model, we find if the sight line is at a fixed angle 
($\theta = -0.05^\circ,$ see, Fig.~\ref{fig(intensity)}) to the
particle trajectory plane, observer tends to receive a single sign circular
polarization. This type of single sign circular polarization has been 
observed in many pulsars (e.g., Han et~al. 1998). As a followup of this 
work, we plan to consider the full 3D dynamics of plasma in a  rotating dipolar
magnetic field, and estimate the coherent radiation.

\section{Conclusion}

By considering projected dipolar magnetic field lines on a plane perpendicular 
to the  rotation axis, we have developed a 2D model for the particle dynamics 
in pulsar magnetosphere. The motive behind developing this  model is to 
elucidate some of the rotational effects induced in the pulsar profiles. We 
have obtained the analytical expressions for the particle trajectory and its 
curvature radius. The energy of particles increase at the expense of neutron 
star's rotational energy. We find the sight line orientation relative to the 
particle trajectory plane might determine the sign of circular polarization. 
The asymmetries observed in the pulse profiles can be explained by considering 
the aberration-retardation effects.
\renewcommand{\theequation}{A-\arabic{equation}}
\setcounter{equation}{0}   
\section*{Appendix-A: To find a series expansion
for the factor $\bf\hat{n}\times( \hat{n} \times \vec{\beta})$ that
appears in Eq.~(\ref{eqefield})}  
Consider $\vec{\beta}$ in Cartesian co-ordinates:
\begin{equation}\label{eqbeta1}
\vec{\bf\beta} =\frac{1}{c}(V_x \hat{x}+V_y \hat{y})\, ,
\end{equation}
where $\hat{x}$ and $\hat{y}$ are the unit vectors along the $x$  and
$y$-axes, respectively, (Fig.~{\ref{fig(caxis)}). Then it follows from 
Eq.~(\ref{eqbeta}) that
\begin{equation}\label{eqvx}
   V_{\rm{x}}=V_{\rm{r}} \cos{\left({\frac{\phi_{\rm{tot}}(t
        +t_{0})}{2}}\right)} - V_{\phi}
        \sin{\left({\frac{\phi_{\rm{tot}}(t+t_{0})}{2}}\right)} \, ,
\end{equation}
and
\begin{equation}\label{eqvy}
 V_{\rm{y}}=V_{\rm{r}}\sin{\left(\frac{\phi_{\rm{tot}}(t+t_{0})}{2}\right)}+
       V_{\phi} \cos{\left(\frac{\phi_{\rm{tot}}(t+t_{0})}{2}\right)}\,  .
\end{equation}
Using Eq.~({\ref{eq_rt0}}), we derive the series expansions for
radial velocity $V_{\rm{r}}$ and rotation velocity $V_{\phi}:$
\begin{equation}\label{eqvr}
      V_{\rm{r}}=\frac{d r}{dt}~,
\end{equation}
\begin{equation}\label{eqvp}
      V_{\phi}=r\Omega^{*} ~.
\end{equation}
By substituting $V_{\rm{r}}$ and  $V_{\rm{\phi}}$ into 
by Eqs.~({\ref{eqvx}}) and ~({\ref{eqvy}}), we obtain
\begin{eqnarray}\label{eq_Vx_Vy_terms}
   V_{\rm{x}} & = & V_{\rm{x0}}+V_{\rm{x1}}t+V_{\rm{x2}} t^2\, ~, \nonumber\\
   V_{\rm{y}} & = & V_{\rm{y0}}+V_{\rm{y1}}t+V_{\rm{y2}}t^2\,  ...\,\,\, .
\end{eqnarray}
The  expressions of $V_{\rm{x0}},\, V_{\rm{y0}},\, V_{\rm{x1}},\,
V_{\rm{y1}}\, ...$ in  the  above expansions are  lengthy (see, Mathew and Gangadhara 2005).
 Using the triple vector identity and the definitions of  
$\hat{n}$, $\hat{\epsilon}_{||}$ and $\hat{\epsilon}_{\perp},$ we obtain
\begin{equation}   
\hat{n}\times( \hat{n} \times \vec{\beta}) = -(\vec{\beta}\, .\,  
       \hat{\epsilon}_{||})\, \hat{\epsilon}_{||}
      -(\vec{\beta}\, .\, \hat{\epsilon}_{\perp})\,
      \hat{\epsilon}_{\perp}\, ,
\end{equation} 
where
\begin{eqnarray}
\vec{\beta}\,. \,\hat{\epsilon}_{||} & = & -\frac{\sin\theta}{c}(V_{\rm{y}}
             \sin\eta + V_{\rm {x}}\,\cos\eta )\, , \\
\vec{\beta}\,.\,\hat{\epsilon}_{\perp} & = & \frac{1}{c}(V_{\rm{y}}\cos\eta- 
     V_{\rm{x}}\sin\eta)\, .
\end{eqnarray}
Using the series expansions of $V_{\rm{x}}$ and $V_{\rm{y}},$ we write
\begin{equation}\label{eq_nnbeta}
\hat{n}\times( \hat{n} \times \vec{\beta})
 =\frac{1}{c}\left[\hat{\epsilon}_{||} \sin{\theta}\, (V_{||\,
 0}+V_{||\, 1}t+V_{||\, 2}t^2) +\hat{\epsilon}_{\perp}(V_{\perp 0}
 +V_{\perp\, 1}t+V_{\perp\, 2})\right]\, ,
\end{equation} 
where
\begin{eqnarray}
 V_{||\,\rm{i}} & = & V_{\rm{yi}}\sin\eta +  V_{\rm{xi}}\cos\eta\, ,\\
 V_{\perp\,\rm{i}} & = & V_{\rm{xi}}\sin\eta - V_{\rm{yi}}\cos\eta\,  ,
\end{eqnarray}
and i = 0, 1, 2\,\,.

\renewcommand{\theequation}{B-\arabic{equation}}
\setcounter{equation}{0} 
\section*{Appendix-B: Transformations for solving Eq.~(\ref{eqefield}) }  
Using the method of Buschauer \& Benford (1976), we make
the following transformations  in order to solve the integral
in Eq.~(\ref{eqefield}).
         
Consider
\begin{equation}\label{eq_exp_trans}
\int\limits_{-\infty}^{\infty}\ {\rm{ exp}}[i(N_1t+N_2t^2+N_3t^3)]\,dt = 
      \frac{1}{N_3^{1/3}} {\rm{e}}^{i C_{\rm n}} 
      \int\limits_{-\infty}^{\infty}\exp[i (z\tau+\tau^3)]\, d\tau\, ,
\end{equation}
where 
\begin{equation}
\tau = \frac{1}{N_3^{1/3}}\left(t+\frac{N_2}{3N_3}\right)
\end{equation}
is a dimensionless variable,  and
\begin{equation}
z = \frac{1}{N_3^{1/3}}\left(N_1-\frac{N_2^{2}}{3N_3}\right)\,.
\end{equation}

By differentiating the Eq.~(\ref{eq_exp_trans}) with respect to
          $N_1$ and $ N_2,$ we obtain
\begin{eqnarray}
\int\limits_{-\infty}^{\infty} t \,\exp[i(N_1t+N_2t^2+N_3t^3)]\, dt = 
   \frac{1}{N_3^{2/3}} {\rm e}^{i C_{\rm n}} \Big[ \int\limits_{-\infty}^{\infty}
   \tau\, \exp[i (z\tau+\tau^3)]\, d\tau \nonumber\\
   -C_l\int\limits_{-\infty}^{\infty} \exp[i(z\tau+\tau^3)]\,d\tau \, \Big]
\end{eqnarray}
and
\begin{eqnarray}          
\int\limits_{-\infty}^{\infty}t^2\exp[i(N_1t+N_2t^2+N_3t^3)]\,dt = 
    \frac{1}{N_3^{1/3}}{\rm e}^{i C_{\rm n}}\Big[C_{\rm m}\int\limits_{-\infty}^{\infty}
    {\rm{exp}}[i(z\tau+\tau^3)]\, d\tau \nonumber\\
    + C_{\rm{p}}\int\limits_{-\infty}^{\infty}\,\tau
    \exp[i(z\tau+\tau^3)]\, d\tau \Big]\, .
\end{eqnarray}
We define
\begin{eqnarray}
{\rm L}_{1}(z) &=& \int\limits_{-\infty}^{\infty}\exp[i(z\tau+\tau^3)]\, d\tau  = 
    \frac{2}{3}\sqrt{z}\,{\rm K}_{1/3} [2(z/3)^{3/2}]\, ,\\
{\rm L}_{2}(z) &=& \int\limits_{-\infty}^{\infty}  \tau \exp[i(\,z\tau+\tau^3)]\, d\tau =i
    \frac{2}{\sqrt{27}}\,z\,{\rm K}_{2/3}[2(z/3)^{3/2}]\, , \\
{\rm B}_0 &=& \frac{1}{N_3^{1/3}}{\rm e}^{i C_{\rm n}}{\rm L}_1(z)\, ,\\
{\rm B}_1 &=& \frac{1}{N_3^{2/3}}{\rm e}^{i C_{\rm n}}[{\rm L}_2(z)-C_{\rm l}{\rm L}_1(z)]\, , \\
{\rm B}_2 &=& \frac{1}{N_3^{1/3}} {\rm e}^{i C_{\rm n}}\left[C_{\rm m}{\rm L}_1(z)-C_{\rm p}
            {\rm L}_2(z)\right]\, ,
\end{eqnarray}
where
\begin{eqnarray}
C_{\rm l} &=& \frac{N_2}{3N_3^{2/3}}\, ,\\
C_{\rm n} &=& \frac{N_2}{3N_3}\left[\frac{2N_2^{2}}{9N_3}-N_1\right]\, ,\\ 
C_{\rm m} &=& \left[\frac{2N_2^2}{9N_3^2}-\frac{N_1}{3N_3}\right]\, ,\\
C_{\rm p} &=& \frac{2N_2}{3N_3^{4/3}}\, .
\end{eqnarray}
  \renewcommand{\theequation}{C-\arabic{equation}}
\setcounter{equation}{0}
 \section*{Appendix-C: The  expressions for  expansion coefficients 
which appear in Eqs.~(\ref{eq_N_0123},~\ref
{eq_Vx_Vy_terms}~) }
  The expression for $\phi_{tot}$  in Eq~.(\ref{eq_phi_tot}) 
is  expanded in a series given by
\begin{eqnarray}
 \phi_{\rm tot}=\phi_0 +\phi_1 \,t +\phi_2\,\, t^2 
+\phi_3 \, t^3.....\,\,\,\nonumber~.
 \end{eqnarray}
\begin{eqnarray}
 \phi_0 &=& \arccos  \mu_3 \nonumber ~,\\
    \phi_1 &=& \Omega  -\frac{\mu_2\,a_1'}{\mu_3}\,
  \left( \frac{{d_0}\,\cos {\phi_p}}{{\mu_1\, a_0'}} -1\right)\nonumber~,\\
  \phi_2 & =& \Big[ {d_0}\,\sin^4\phi_p\,
    \Big( {{d_0}}^4\,\Big\{ \cos (4\,{\phi_p})
  -\cos (2\,{\phi_p}) \Big\} \nonumber\\ & +&
      {a_0}'\,\Big[ {{d_0}}^2\,
          \Big\{ 1 + 5\,\cos (2\,{\phi_p}) \Big\} \,{a_0}' +
         2\,{{a_0}'}^3 - 2\,{d_0}\,{\mu_1}\,
          \Big\{ {{d_0}}^2\,\cos (3\,{\phi_p}) \nonumber\\ & + &
            3\,\cos {\phi_p}\,{{a_0}'}^2 \Big\}  \Big]  \Big) \,
    \Big\{ -2\,\,{{a_1}'}^2 +
      4\,\,{a_0}'\,{a_2}' \Big\}\Big]\frac{1 }{8\,
    {{\mu_3}}^2\,{{\mu_4}}\,{{a_0}'}^5}  \nonumber~,\\
   \phi_3 &= & \frac{\mu_{10}}{3} -
    \frac{1}{\sqrt{1 - {\mu_3}^2}}
 \Big( \frac{{\mu_1}\, {\mu_{13}}\,\cos {\phi_p}}{ 3}
 + \frac{{d_0}\,{\mu_{12}}\,
            \sin^2{\phi_p} }{{a_0}'}
         \Big) \nonumber~.
\end{eqnarray}

    The coefficients $ N_0, N_1,N_2\,\, {\rm and}\,\, N_3$ which appear
in the series expansion of Eq.~(\ref{eq_N_0123}) are given  by 
   \begin{eqnarray}
 N_0 & = & \omega \,\left( {t_0} -
    \frac{\cos \theta \,
       \cos (\eta  - {\phi_0})\,{a_0}'}{
       c} \right)~, \\
 N_1 & = & \frac{\omega}{c} \,\left[ c -
      \cos \theta \,\left\{ {\phi_1}\,
          \sin (\eta  - {\phi_0})\,{a_0}'
          + \cos (\eta  - {\phi_0})\,
          {a_1}' \right\}  \right] ~,\\
N_2 &= &\frac{\omega}{2\, c} \,\cos \theta \,
   \Big[ \{ {{\phi_1}}^2\,
          \cos (\eta  - {\phi_0}) -
         2\,{\phi_2}\,
          \sin (\eta  - {\phi_0}) \} \,
       {a_0}'   \nonumber\\ & - &
 2\,\{ {\phi_1}\,
          \sin (\eta  - {\phi_0})\,{a_1}'
          + \cos (\eta  - {\phi_0})\,
          {a_2}' \}  \Big]   ~,\\
 N_3 &= &\frac{ \omega}{6\,c} \,\cos \theta \,
    \Big[ \Big\{ 6\,{\phi_1}\,
          {\phi_2}\,
          \cos (\eta  - {\phi_0}) +
         ( {{\phi_1}}^3 -
            6\,{\phi_3} ) \,
          \sin (\eta  - {\phi_0}) \Big\} \,
       {a_0}'   \nonumber \\
& + &   3\,\Big\{ {{\phi_1}}^2\,
          \cos (\eta  - {\phi_0}) -
         2\,{\phi_2}\,
          \sin (\eta  - {\phi_0}) ) \,\Big\}
       {a_1}' \nonumber \\ & -  &
      6\,\Big\{ {\phi_1}\,
          \sin (\eta  - {\phi_0})\,{a_2}'
          + \cos (\eta  - {\phi_0})\,
          {a_3}' \Big\} \Big] ~.
\end{eqnarray}
 The terms appearing in   the series expansion in 
(\ref{eq_Vx_Vy_terms}) are  given by  
   \begin{eqnarray}
 V_{\rm x0} & = & -\left( \Omega \,\sin {\phi_0}\,
 {a_0}'\,\,{\sqrt{{{a_0}'}^2 - {{{b}}^2}}}\,
      \right)  + \cos {\phi_0}\,{a_1}' ~, \\
   V_{\rm x1} & = & - {\phi_1}\,\sin {\phi_0}\,{a_1}'+
 2\,\cos {\phi_0}\,{a_2}'  \nonumber\\
 & + & \Big[{\phi_1}\,\Omega \,\cos {\phi_0}\,
      \left( {{b}}^2 - {{a_0}'}^2 \right)  -
     \Omega \,\sin {\phi_0}\,{a_0}'\,{a_1}'\Big]
    \Big({{{a_0}'}^2 - {{b}}^2 }\Big)^{-1/2} ~, \\
V_{\rm x2} & =  -& \left[  {{\phi_1}}^2\,\cos {\phi_0}
      + 2\,{\phi_2}\,\sin {\phi_0} \right] \,{a_1}' -
    4\,{\phi_1}\,\sin {\phi_0}\,{a_2}'
+6\,\cos {\phi_0}\,{a_3}'  \nonumber \\ &+&
    \Big[2\,\Omega \,\cos {\phi_0}\,
        ( {b} - {a_0}' ) \,
        ( {b} + {a_0}' ) \,
        \{({{a_0}'}^2 - {{b}}^2\,){\phi_2}   +
           + {\phi_1}\,{a_0}'\,{a_1}' \}    \nonumber \\ &+ &
       \Omega \,\sin {\phi_0}\,
        \{ {{b}}^4\,{{\phi_1}}^2 -
          2\,b^2\,{{\phi_1}}^2\,{{a_0}'}^2 +
          {{\phi_1}}^2\,{{a_0}'}^4 +
   \nonumber \\ & +&
          {{b}}^2\,{{a_1}'}^2 +
          2\,{a_0}'\,( b^2 - {{a_0}'}^2 ) \,{a_2}'
          \}\Big] \Big[{2({{a_0}'}^2 - {b^2})^{3/2}}\,
 \Big]^{-1}  ~,\\
 V_{\rm y0} &=& \Omega \,\cos {\phi_0}\,
{\sqrt{{{a_0}'}^2 - {b^2}}}\,
    + \sin {\phi_0}\,{a_1}'~,\\
 V_{\rm y1} & = & {\phi_1}\,\cos {\phi_0}\,{a_1}'+2\,\sin {\phi_0}\,{a_2}'
 \nonumber\\
 &+& \Big[{ {\phi_1}\,\Omega \,\sin {\phi_0}\,
      \left( {{b}}^2 - {{a_0}'}^2 \right)  +
     \Omega \,\cos {\phi_0}\,{a_0}'\,{a_1}'\Big]
     \Big({{{a_0}'}^2 -
 {b^2}}\,}\Big)^{-1/2} ~, \\
 V_{\rm y2} &  = &  3\,\sin {\phi_0}\,{a_3}'+ \Big( {\phi_2}\,\cos {\phi_0} -
     \frac{{{\phi_1}}^2\,\sin {\phi_0}}{2} \Big) \,{a_1}'
   + 2\,{\phi_1}\,\cos {\phi_0}\,{a_2}' \nonumber\\ &+ &
 \Big[ \Omega \,\Big\{- \Big(  {{\phi_1}}^2\,
             \cos {\phi_0}  +
          2\,{\phi_2}\,\sin {\phi_0} \Big) \,
        {\Big( b^2 - {{a_0}'}^2 \Big) }^2 \nonumber\\
 &+& 2\,{\phi_1}\,\sin {\phi_0}\,{a_0}'\,
        \Big( {{b}}^2 - {{a_0}'}^2 \Big) \,{a_1}' -
       b^2\,\cos {\phi_0}\,{{a_1}'}^2 \nonumber\\ &+&
       2\,\cos {\phi_0}\,{a_0}'\,
        \Big({{a_0}'}^2 -{{b}}^2 \Big) \,{a_2}' \Big\}\Big]
  \Big[{2({{a_0}'}^2 - {b^2})^{3/2}}\,
 \Big]^{-1}~,
\end{eqnarray}
    \begin{eqnarray}
 \mu_1 &=& {\sqrt{1 - \frac{{{d_0}}^2\,
{\sin^2 \phi_p}}{{{a_0}'}^2}}}  \nonumber~,\\
 \mu_2 &=&  \frac{{d_0}\,{\sin^2 \phi_p}}{{a_0}'}\nonumber~,\\
 \mu_3 &=& {\sqrt{1 - {\left( {\mu_2} + {\mu_1}\,\cos {\phi_p} \right)
         }^2}} \nonumber~,\\
  \mu_4 & =&  {{a_0}'}^2\,\,{\mu_1}^2\nonumber~,\\
 \mu_5 &=& \frac{{\mu_2}^2\,\cos\phi_p\,a_1'}{\mu_1\,a_0'\,
\sin^2\phi_p}\nonumber~,\\
 \mu_6 &=&\Big(\frac{\mu_2^2\,a_1'\,}
{\mu_1^2\,a_0\,\sin^2\phi_p}\Big)^2\nonumber~,\\
  \mu_7 &=&
      (\frac{\mu_2}{\mu_1\,\sin\phi_p})^2
\left(\frac{2\,a_2'}{a_0'}-\frac{3\,a_1^2}{a_0'^2}\right)
-\mu_6  \nonumber~,\\
  \mu_8 &=& \frac{\mu_1\,\mu_7\,\cos\phi_p}{2}
  +\mu_2\left(\frac{a_1'^2}{a_0'^2}-\frac{a_2'}{a_0'}\right) \nonumber~,\\
  \mu_9 &=& \Big\{ \left( 1 + 2\,{{\mu_3}}^2 \right) \,
     {{\mu_5}}^2 -
    2\,{\mu_3}\,
     \left(  {{\mu_3}}^2-1 \right) \,
     {\mu_8} \nonumber\\ & + &
    \Big\{{d_0}\,
       \left( 1 + 2\,{{\mu_3}}^2 \right) \,
       \sin^2 {\phi_p}\,{a_1}'\,
       \left( \mu_2\,a_1'-2\,{\mu_5}\,{{a_0}'}
           \right)\Big\}\frac{1}{ {{a_0}'}^3}\Big\}\frac{1}
    {\left(  {{\mu_3}}^2-1 \right)^2 }  \nonumber~,\\
  \mu_{10} &=& \frac{\left( -4\,{\mu_3}\,{\mu_8} +
      \left( -1 + {{\mu_3}}^2 \right) \,
       {\mu_9} \right) \,
    \left( {\mu_5}\,{{a_0}'}^2 -
      {d_0}\,\sin^2 {\phi_p}\,
       {a_1}' \right) }{2\,
    {\left( 1 - {{\mu_3}}^2 \right) }^
     {\frac{3}{2}}\,{{a_0}'}^2}  \nonumber~,\\
  \mu_{11} &=& \frac{-6\,\left( 2\,{{a_1}'}^3 -
      3\,{a_0}'\,{a_1}'\,
       {a_2}' +
      {{a_0}'}^2\,{a_3}' \right) }{{
       {a_0}'}^3}  \nonumber~,\\
 \mu_{12} &=&
-\left( \frac{{{a_1}'}^3 -
      2\,{a_0}'\,{a_1}'\,
       {a_2}' +
      {{a_0}'}^2\,{a_3}'}{{{a_0}
        '}^3} \right)   \nonumber~,\\
\mu_{13} &=&- \frac{{\mu_2}^2\, \left( {\mu_{11}}\,{a_0}' +
      3\,{\mu_7}\,{a_1}' \right) }
{\mu_1^2\,\,2\,{{a_0}'} \,\sin^2{\phi_p}}\nonumber~.
\end{eqnarray}
\begin{acknowledgement}
We thank Jayanth Murthy for comments. 
\end{acknowledgement}

\end{document}